\documentclass[10pt]{iopart}

\usepackage{iopams}  
\begin{document}

\title[Schr\"{o}der's functional equation]{The Schr\"{o}der functional equation and its relation to the invariant measures of chaotic maps\footnote{ This is an author-created, un-copyedited version of an
  article accepted for publication in J. Phys. A: Math. Theor. 41 (2008)
  265101. IOP Publishing Ltd is not responsible for any errors or omissions in
  this version of the manuscript or any version derived from it. The definitive
  publisher authenticated version is available online at
doi:  10.1088/1751-8113/41/26/265101}}

\author{Jos\'e-Rub\'en Lu\'evano$^1$, Eduardo Pi\~na$^2$} 
\address{$^1$ Departamento de Ciencias B\'asicas, Universidad Aut\'onoma
  Metropolitana, Unidad Azcapotzalco,\\ M\'exico, D.F., CP 02200, M\'EXICO}
\ead{jrle@correo.azc.uam.mx}  

\address{$^2$ Departamento de F\'isica, Universidad Aut\'onoma
  Metropolitana, Unidad Iztapalapa,\\ M\'exico, D.F., CP 09340, M\'EXICO}
\ead{pge@xanum.uam.mx}

\begin{abstract}
The aim of this paper is to show that the invariant measure for a class of one dimensional chaotic maps, $T(x)$, is an extended solution of the Schr\"{o}der functional equation, $q(T(x))=\lambda q(x)$, induced by them. Hence, we give an unified treatment of a collection of exactly solved examples worked out in the current literature. In particular, we show that these examples belongs to a class of functions introduced by Mira, (see text). Moreover, as a new example, we compute the invariant densities for a class of rational maps having the Weierstrass $\wp$ functions as an invariant one. Also, we study the relation between that equation and the well known Frobenius-Perron and Koopman's operators.
\end{abstract}
\date{\today}

\noindent

\maketitle

\section{Introduction}
The notion of an invariant measure plays a central role in the statistical characterization of chaotic dynamical systems. In spite of the well established theorems about the existence of such measures for abstract dynamical systems the construction of such measure for a given dynamical system is in general a difficult task. In the case of chaotic maps it is well known that if the invariant measure has a density, this one is a solution of the Frobenius-Perron functional equation. In one dimensional dynamics, some rigorous results about the existence of such measures are known, see for example \cite{bj,ce,gr,kch,kh,l,lmc}.
There is not doubt about the  physical significance of the invariant measures, with them we can compute \emph{ensambles} averages, correlations functions, etc., in particular modeling non-equilibrium states \cite{bs}. Therefore to find reliable analytical methods to construct these measures has special interest.

In the last three decades a series of papers, devoted to deterministic chaos, are reporting examples of chaotic maps that are handled by means of a change of variables method \cite{bo,ga,gr,gth,kf,kh,lmc,m,mr,sg,sr,u1}. Their study concerns with nonlinear piecewise transformations that are topologically conjugated, or semi-conjugated, to linear piecewise maps. Hence, the invariant density for the nonlinear case is obtained by transforming the constant piecewise density of the linear one. In particular, it has been remarked in some of these papers that the invariant measure of the nonlinear map is in fact the function doing such conjugation. Also, this procedure allows to generalize some results proved for piecewise linear maps to nonlinear maps conjugated to them. For example, existence theorems, ergodic and asymptotic properties or spectral decompositions of nonlinear transformations \cite{bs,gth,lmc}.    

Moreover, from the experimental point of view, one-dimensional maps are useful to modeling nonlinear processes in diverse fields of science. Also, from theoretical point of view with the study of nonlinear maps, which are conjugated to piecewise maps, we gain insight to understand integrable chaos \cite{d,m}. 

In this paper we provide a new approach to the problem of computing invariant densities and measures for chaotic maps, based on Schr\"oder's functional equation. This method becomes an alternative to the traditional one given by the Frobenius-Perron equation, but also more important, that can be interpreted as an adjoint problem to this one. The main difference between these two approaches comes from the fact that, for Schr\"oder's equation we only need to know the transformation and its derivative, without the explicit knowledge of the inverse transformation and its respective derivative, as it is demanded for the Frobenius-Perron equation. Also, it is very interesting to us the fact that Schr\"oder's equation is older than the Frobenius-Perron equation. Then, we are giving a new look to an old equation. 
 
	This paper is divided in seven parts, section II is concerning with a brief exposition of the well known conjugation property between maps. The section III is presenting several exactly solvable examples. The fourth one, presenting the main result: the formal relation between the Schr\"{o}der and Frobenius-Perron equations, hence, we proceed to the construction of the invariant density. In section V we give a new example, a class of functions having the Weierstrass $\wp(x)$ function as an invariant. Finally, we compute the Lyapunov's exponent of these maps and discuss some possible generalizations of our method.

\section{Conjugate maps}
Let us consider a transformation $T$ of an interval $I$ into itself,
which preserves the measure $\mu$, i.e. $\mu(A)= \mu(T^{-1}(A))$. Let $x \in I$, the iteration 
\begin{equation}
 x_{n+1}=T(x_n)\; ,
\label{iter}
\end{equation}
defines a discrete dynamical system.
Now, if there is a transformation $S:I\to I$ related to $T$ by a change of variables
$T(h(x))=h(S(x))$, where $h:I \to I$ is a continuous one-to-one
function, we say that the maps $T$ and $S$ are topologically conjugate. If in addition, $h$ is a piecewise differentiable function then their invariant densities are related by:
\begin{equation}
  \label{pp}
  \rho_{T}(x)= \frac{\rho_{S}(h^{-1}(x))}{\left| h^{\prime}\circ h^{-1}(x)\right|}\;.
\end{equation}
The simplest examples of linear piecewise transformations of an interval are: the \emph{R\'enyi} transformation $R_{r}(x):=\{rx\}$ and the piecewise continuous transformations \cite{gr,gth,lmc}
\begin{equation}
N_{r}(x):=(-1)^{[rx]}\{rx\}\;,
\end{equation}
where $[rx]$ is the integer part of $rx$ and $\{rx\}$ means $rx\bmod 1 = rx-n$, where $n$ is the largest integer such that $rx-n\geq 0$ for any $r \in \mathbb{N}$.

In the following we are concerning with the case when $S$ is $R_{r}$
or $N_{r}$, hence $\rho_{S}(x)=1_{[0,1]}(x)$, for any $r\in \mathbb{N}$. Therefore, $\rho_{T}(x)= 1/\left| h^{\prime}\circ h^{-1}(x)\right|$. Also, we denote by $T_{r}$ any map conjugated to $N_{r}$ or $R_{r}$. It is implied by the conjugation property that each map $T_{r}$ has $r$ monotonic pieces that map the interval $I$ onto itself. Such that, the inverse function $T^{-1}$ has $r$ monotonic branches on $I$, each one denoted by $T_{i}^{-1}$. 

\section{Exactly solvable maps}
To show the essence of our method we work out in detail the logistic transformation. It is well known that it is conjugated to $N_2$ and that its invariant density satisfies the Frobenius-Perron equation. 

\subsection{Logistic map}\label{logmap}
 Let us consider the transformation $F(x)=4x(1-x)$, where $x\in [0,1]$. The function $h(\theta)=sin^2(\frac{\pi}{2}\theta)$, $\theta
\in [0,1]$ satisfies:
\begin{equation}\label{conj}
F[h(\theta)]=h[N_{2}(\theta)],
\end{equation}
hence, by \eref{pp}
\begin{equation}
\rho_{F}(x)=\frac{1}{\pi \sqrt{x(1-x)}}.
\end{equation}
Finally, the invariant measure for $F$ is 
\begin{eqnarray}
\mu(x)&=& \int_{0}^{x}\frac{1}{\pi \sqrt{y(1-y)}}dy , \qquad x \in I ,\\
&=& \frac{2}{\pi}\arcsin(\sqrt{x})\;,
\end{eqnarray}
in other words, $\mu(x)=h^{-1}(x)$. It should be noted that \eref{conj} can be written as $F(x)=h[N_{2}(h^{-1}(x))]$ then, by periodicity of the $sin$ function the last expression is equal to $F(x)=h(2h^{-1}(x))$, i.e. $F(x)=\sin^2(2\sin^{-1}\sqrt{x})$.
 
\subsection{A collection of exactly solvable maps}\label{exact} 
Exactly solvable iterations are known since the nineteenth century, the interest was to understand Newton's method \cite{bo,cg,mil}. However, in the last thirty years a boom on the search of analytic results for chaotic systems shifted the attention to the topic.

 A very important examples are provided by the Chebyshev polynomials. It is well known that they are ergodic and mixing maps \cite{lmc}. From its definition: $T_{r}(x)=\cos(r\cos^{-1}x)$, we can see that they are conjugated to $N_{r}(x)$ maps. To show this, we start with the function $h(\theta)=\cos(\pi\theta)$, and obtaining
$T_{r}(x)= h[N_{r}(h^{-1}(x))]$. Hence, using \eref{pp}
$\rho_{T}(x)=1/\pi \sqrt{1-x^2}$. Such that, its invariant measure is $\mu(x)=\frac{1}{\pi}\arccos(x)$. We can see also that $F(x)$ is also conjugated to $T_{2}(x)$ \emph{via} $N_{2}(x)$.

Now, we list some papers on the subject: a) in \cite{gr} the invariant density of maps conjugated with the Jacobian elliptic sine function, $sn(x)$, are presented. The class of maps studied is $T_{r}(x)=f(sn(rsn^{-1}(f^{-1}(x))))$, where $f(x)$ is an appropriate function mapping the interval $[0,1]$ onto itself; b) in \cite{kf} some examples of one dimensional iterations are solved using the addition properties of trigonometric and elliptic functions; c) in \cite{ga} transformations of the real line associated to special statistical distributions are built. In particular, the Cauchy, $F$ and $Z$ distributions are associated to maps of the Real Line conjugated to $N_{2}$; d) in \cite{bo,sr} Newton's algorithm to search the square root of $\sqrt{-3}$ is found to be conjugated to $N_{2}$, or as in \cite{sg} to $2x+1/2 \bmod 1$; e) recently, in \cite{u1} a similar method as \cite{gr} is given; f) finally, an exception is the paper \cite{mr} where Schr\"oder's equation (see below) is explicitly used in order to generalize the properties of the Chebyshev polynomials to more general functions. But, in that paper the relation to the Frobenius-Perron equation is only suggested. Nevertheless, it should be noted that the invariant densities of all those examples are exactly computed exploiting the addition and periodicity properties of trigonometric and Jacobian elliptic functions jointly with the conjugation property with $R_{r}$ or $N_{r}$. It is worth to mention, that the examples handled in these papers were obtained by each author mostly in an independent way to the others ones. For that reason we speak about \emph{rediscovered} maps in the last paragraph of \Sref{cinco}.

\section{Schr\"oder's equation}
 By definition, if $\mu(x)$ has a density $\rho(x)$, then 
\begin{equation}
\mu(x)= \int_{0}^{x}\rho(y)dy \;, \qquad x \in I\; .
\end{equation}
Now, the preservation of $\mu$ under $T$ implies that $\rho(x)$ satisfies the \emph{Frobenius-Perron} functional
equation \cite{ce}:
\begin{equation}
\rho(x) = \sum_{i=1}^{r}\frac{\rho(T_{i}^{-1}(x))}{\left|
  T^{\prime}\circ T_{i}^{-1}(x)\right|} \;,
\end{equation}
or, in an integral form:
\begin{equation}
\rho(x)=\int_{I}\delta(x-T(y))\rho(y)dy\;.
\end{equation}
	Now, using this last expression, we ask the question about the left eigenvalue problem for the Frobenius-Perron
equation, which is defined by:
\[
\lambda q(x)=\int_{I} q(y)\delta(y-T(x))dy \; = \; q(T(x))\;.
\]
This equation is nothing but Schr\"oder's functional
equation \cite{cg,kcz}: 
\begin{equation}
\lambda q(x)=q(T(x))\;, 
 \label{eqsch}
 \end{equation}
where $\lambda \in \mathbb{R}$ and $q(x)$ is a function of a real variable.
Hence, the iteration \eref{iter} is linearized, such that its  $n$-th
iterate is (at least locally): $\lambda^n q(x) = q(T^n(x))$.

	The classical result in the local analysis of solutions of
Schr\"oder's equation is Koenig's theorem
which guarantees the existence of a solution in the neighborhood of a
fixed stable point of $T$ \cite{kch,kcz}. However, more difficult is our present interest in the global study of that equation, see \cite{bj,kch,kcz} for a discussion. 

	Now we will consider a differentiable map $T:I \to I$ conjugated to $N_{r}(x)$, then $T$ has
$r$ monotonic pieces on $I$, and $T^{-1}$ has $r$ branches on $I$. We
proceed as follows: taking the derivative on both sides of Schr\"oder's equation \eref{eqsch} and 
after taking its absolute value, we obtain
 \begin{equation} \label{eqd}
  \left| \lambda \right| \left| q^{\prime}(x)\right| = \left| T^{\prime}(x)\right|
  \left| q^{\prime}(T(x))\right| \;.
 \end{equation}
Indeed, substituting the $j$-th branch of $T^{-1}$ in \eref{eqd}
 \begin{equation}
  \left| \lambda \right| \left| q^{\prime}(T_{j}^{-1}(x))\right| = \left| T^{\prime}(T_{j}^{-1}(x))\right|
  \left| q^{\prime}(x)\right|\;.
 \label{nte3}
 \end{equation}
$x \in I$.
We observe that this expression is valid for any $j=1,\ldots,r$.
We are now ready to introduce the result of this paper, putting together all terms  
of the form \eref{nte3}:
 \begin{equation}
 \left| \lambda \right| \sum_{j=1}^{r}\left|\frac{
 q^{\prime}(T_{j}^{-1}(x))}{ T^{\prime}(T_{j}^{-1}(x))}\right| =
 r \left| q^{\prime}(x)\right| \;,
  \label{esti}
 \end{equation}
where $j$ runs over all the inverse branches of $T$. As we can see,
this is the Frobenius-Perron equation for $\left| q^{\prime}(x) \right|$, therefore, if there exist a function $\alpha(x) = \left| q^{\prime}(x) \right|$ as a global solution of \eref{esti}, defined on $I$,
for $\left| \lambda \right|\, = r$, then the corresponding invariant measure
of $T$ is
\begin{equation}\label{meas}
\mu(x)= \int_{0}^{x}\left| q^{\prime}(y)\right| dy\;, \qquad y \in I\;.
\end{equation}
	It is very easy to prove, by direct substitution, that the
examples handled in \cite{bo,ga,gr,gth,kf,lmc,mr,sg,sr,u1}, which are conjugated to $N_{r}$, satisfies the equation \eref{eqd}.
For example, we use the logistic map to show the method. First, substituting $F_{\pm}^{-1}(x)=(1\pm \sqrt{1-x})/2$ in the lhs of \eref{eqd},
\begin{equation}	
\left|\lambda\right|\rho(F_{\pm}^{-1}(x))=\frac{\left|\lambda\right|}{\pi\sqrt{F_{\pm}^{-1}(x)(1-F_{\pm}^{-1}(x))}}=\frac{2\left|\lambda\right|}{\pi\sqrt{x}}\;,
\end{equation}
and in rhs of \eref{eqd}, 
\begin{equation}	
\left|F^{\prime}(F_{\pm}^{-1}(x))\right|\rho(x)=\frac{4\sqrt{1-x}}{\pi\sqrt{x(1-x)}}=\frac{4}{\pi\sqrt{x}}\;,
\end{equation}
such that, we have an equality if and only if $\left|\lambda\right|=2$. 

	It is worth mention that the \eref{eqd} is formally different
from \eref{eqsch}. However, if the functional equation
$\alpha(T(x))=\frac{\lambda}{T^{\prime}(x)}\alpha(x)$ has a continuous
solution $\alpha(x)$ in $I$, fulfilling $\lambda =
T^{\prime}(\overline{x})$, then $q(x)=\int_{0}^{x}\alpha(x)dx$ is a $C^{1}$ solution of
\eref{eqsch}. Thus, there exists a one-to-one correspondence between
the $C^{1}$ solutions of Schr\"oder's equation and the continuous
solutions of \eref{eqd} \cite{kcz}.

\section{Rational transformations on the Real Line}\label{cinco}
 Now, as a non trivial example we consider a class of rational transformations having the Weierstrass $\wp(z)$ elliptic function as an invariant one. This function is defined by an infinite series, also it is known its duplication formula. 
 
  The Weierstrass $\wp(z)$ elliptic function is defined as \cite{as}:
\[
\wp(z)= \frac{1}{z^2}+\sum_{m,n}
\frac{1}{(z-2m\omega_{1}-2n\omega_{2})-(2m\omega_{1}+2n\omega_{2})}\;,
\]
where $z \in \mathbb{C}$ and $\omega_{1}$, $\omega_{2}$ are two numbers the ratio of which
is not real, and the summation is take over all $m,n \in
\mathbb{Z}$ but excepting the case when $m=n=0$ simultaneously. This
is an example of a double periodic function with fundamental periods
$2\omega_{1}$ and $2\omega_{2}$. As it is standard, that function is parametrized with two numbers, $g_{2}$ and $g_{3}$ which are called the elliptic
invariants (and are functions of $\omega_{1},\omega_{2}$). We describe Schr\"oder's method to obtain a rational function invariant under $\wp(x)$: there is a duplication formula \cite{as}:
$$\wp(2z)=-2\wp(z)+\left[\frac{\wp^{\prime\prime}(z)}{2\wp^{\prime}(z)}\right]^2\;,$$
such that, using the identities:
$\wp^{\prime\prime}(z)=6\wp^2(z)-\frac{g_{2}}{2}$, and
$(\wp^{\prime})^{2}(z)=4\wp^3(z)-g_{2}\wp(z)-g_{3}$, we can built a two parametric rational function
\begin{equation}\label{tw}
T_{g_{2},g_{3}}(z)= \frac{z^4 + \frac{g_{2}}{2}z^2+ 2g_{3}z + (\frac{g_{2}}{4})^2}{4z^3-g_{2}z-g_{3}}\;,
\end{equation}
having $\wp(z)$ as an invariant function, i.e. $\wp(2z) = T_{g_{2},g_{3}}(\wp(z))$. Now, let us consider the case
$g_{2},g_{3} \in \mathbb{R}$, then $\wp$ takes real values on the real
line. The inverse function for $\wp$ is also known \cite{a}:
$$\wp^{-1}(u)=\int_{u}^{\infty}\frac{ds}{\sqrt{4s^3-g_{2}s-g_{3}}}
\;.$$ Now, if the \emph{discriminant}
$\Delta=\frac{1}{16}(g_{2}^3-g_{3}^3)$ is positive, then
$4s^3-g_{2}s-g_{3}=0$ has three real and distinct roots and $\wp$ has
the period $\omega_{1}\in \mathbb{R}$.
 Also, the above relations are unaltered by the choice $z = x$. Now, to use the Frobenius-Perron equation we must compute the inverse function of the $T$ given by \eref{tw}, and its derivative. Instead of this, we proceed to compute the solution of the corresponding Schr\"oder's equation associated to such $T$. We can see that in analogy with \sref{logmap} that $\wp(2\wp^{-1}(x))=T_{g_{2},g_{3}}(x)$, therefore the invariant density is given by
\begin{equation}\label{idw} 
\rho(x)=\left| \frac{d}{dx} \wp^{-1}(x)\right|=\frac{1}{\sqrt{4x^3 - g_{2}x - g_{3}}}\,,
\end{equation}
i.e., the eigenfunction of \eref{eqsch} induced by $T_{g_{2},g_{3}}$
and having eigenvalue $\left| \lambda \right| = 2$. 
  
For example, if $g_2=4$ and $g_3=0$, we have the rational transformation $T(x)=(x^2+1)^2/4x(x^2-1)$, where $x\neq 0,\pm 1$, then $T(\wp(x))=\wp(2x)$. This example was given by Latt\`es in 1918 \cite{cg}. Another important example, is given by 
\[
T(x)=\frac{4x(1-x)(1-\kappa^{2}x)}{(1-\kappa^{2}x^{2})^2}\;,
\]
which satisfies the semiconjugacy relation $T(sn^2(x))=sn^2(2x)$, where $sn(x)$ is the Jacobi elliptic sine function. As it is remarked by Milnor \cite{mil}, this example given also by Latt\`es, was in fact studied by Schr\"oder in 1871 \cite{sch}. However, we found that it was also rediscovered in \cite{gr,kf,u1}. A different example, studied by Boole in 1872, is the linear fractional transformation
\[
T(x)=\frac{ax+b}{cx+d}, \; ad-bc=1\;, \; c\neq 0\;,
\]
whose Schr\"oder's equation is solved using an iterative method, called by him as Laplace's method, which it is exposed in \cite{b}. For this example, our method gives $\rho(x)=1/(1+x^2)$ as its invariant density, provided $|a+d|>2$. Again, this is a case of a rediscovered example \cite{sr}.

\section{Lyapunov's exponents}
Finally, the Lyapunov exponent $\Lambda(T)$ of the maps $T_{r}$ is computed using
Schr\"oder's equation induced by it. By definition, $\Lambda(T):=\int_{I}\ln \left| T_{r}^{\prime}(x)\right| \rho(x)dx$, then taking logarithm of \eref{eqd}, and using the invariance of $\mu$ under $T_{r}$, we have from the previous equation the expected result $\Lambda(T)=\ln r$. In other words, $\Lambda(T)$ is equal to the logarithm of the number of monotonic pieces of $T_{r}(x)$.

\section{Discussion and Conclusions}
To the best of our knowledge, we are the first to point out the significance of Schr\"oder's equation to compute invariant densities and measures for chaotic maps. Our approach makes the main difference with the corresponding work of the authors in \cite{bo,ga,gr,gth,kf,kh,l,lmc,m,mr,sg,sr,u1}. As it is mentioned in \Sref{exact}, the work by Mira \cite{mr} using Schr\"oder's equation, is focused in the generalization of the properties of Chebysev maps. It should be noted the interest of authors working on the subject of functional equations on the problem of computing invariant densities, see for example \cite{bj,kch,l}. On the other hand, authors working in dynamical chaos are interested in methods to solve the Frobenius Perron functional equation \cite{bs,bo,ce,d,ga,gr,gth,kf,kh,l,lmc,m,mr,sg,sr,u1}. However, in the current literature the problem of computing invariant densities or measures appears unrelated to the problem of solving Schr\"oder's functional equation.
 
 We remark that the measure $\mu(x)$, which is the integral of $|q^{\prime}(x)|$ in \eref{meas}, could be not equal to $q(x)$, is in that sense that we speak about an \emph{extended} solution of Schr\"oder's equation.
 
Also, we note that in Ergodic Theory Koopman's operator is introduced as the formal adjoint of the Frobenius-Perron operator on densities in the $L^{\infty}$ sense \cite{ce,lmc,m}. In our context, we are considering Schr\"oder's equation as a left eigenvalue problem associated to the Frobenius-Perron functional equation, such that our case is a more general one. It should be pointed out that this investigation is not limited to the one dimensional case. However, higher dimensional dynamics seems to be more difficult and may be requiring a different approach, mainly due to the formal definition of Schr\"oder's equation in several variables \cite{cmc}. 

In this paper we are mainly concerned with real transformations on the interval, but another interesting aspect of the subject is the study of the complex Schr\"oder's equation associated to chaotic transformations of the complex plane \cite{bsh,cg,cmc,sg}. 

Finally, in the one dimensional case becomes very interesting the study of strange attractors at $2^{n}$ banded chaos of the logistic map $F_a(x)=ax(1-x)$. Here $a\in (a_{\infty},4)$, and $a_{\infty}$ is the accumulation point of the period doubling bifurcation \cite{gth,lmc,m}. Further studies in this direction will be published elsewhere. 

\section*{Acknowledgments}
We wish to thank Dr. Ismael Mu\~noz M., Dr. Jos\'e Luis del R\'io C. and Dr. Jaime Cruz Sampedro for valuable comments. This work has been partially supported by CONACYT, SNI-35793, M\'exico.


\section*{References}
\begin{thebibliography}{99}

\bibitem{as} Abramowitz M and Stegun I A (eds) 1970 {\it Handbook of
    Mathematical Functions}, National Bureau of Standards, AMS 55

\bibitem{ac} Acz\'el J 1966 {\it Lectures on Functional Equations and their Applications} Mathematics in Science and Engineering {\bf 19} (New York: Academic Press Inc.)  

\bibitem{a} Ahlfors L V  1979 {\it Complex Analysis} (McGraw-Hill
  3rd. Ed.)

\bibitem{bj} Baron K and Jarczyk W 2001 Recent results on functional equations in a single variable, perspective and open problems (Survey Paper) {\it Aequationes Math.} {\bf 61} 1-48

\bibitem{bs} Beck C and Schl\"ogl F 1993 {\it Thermodynamic of chaotic systems} (Cambridge: Cambridge University Press)

\bibitem{bo} Bender C M and Orszag S A 1978 {\it Advanced
    Mathematical Methods for Scientists and Engineers} (New York: McGraw-Hill)
    p~233-39 
    
\bibitem{b} Boole G 1872 {\it A Treatise on the Calculus of Finite Differences.} (New York: Chelsea Publishing Co., Fourth Ed., Reprinted by Dover) 

\bibitem{bsh} Bourdon P S and Shapiro J H 1997 Mean Growth of Koenigs Eigenfunctions {\it Jour. Amer. Math. Soc.} {\bf 10}(2) 299-325

\bibitem{cg} Carlesson L and Gamelin T W 1993 {\it Complex Dynamics}
  (New York: Springer-Verlag) 

\bibitem{cr} Castillo E and Reyes Ruiz-Cabo M 1992 {\it Functional Equations and Modeling in Science and Engineering} (New York: Pure and Applied Math, Marcel Dekker, Inc.) 

\bibitem{ce} Collet P and Eckmann J P 1980 {\it Iterated Maps on the Interval as Dynamical
  Systems} (Boston: Birkh\"auser)

\bibitem{cmc} Cowen C C and MacCluer B D 2003 Schr\"oder's Equation in Several Variables {\it Taiwanese J. Math.} {\bf 7}(1) 129-54

\bibitem{d} Dorfman J R 1999 {\it An introduction to chaos in nonequilibrium statistical mechanics}, Cambdrige Lect. Notes in Phys. (Cambridge: Cambridge University Press)

\bibitem{ga} Golubentsev A F and Anikin V M 1998 The Explicit Solutions of the Frobenius-Perron Equation for the Chaotic Infinite Maps {\it Int. J. Chaos Bifurc.} {\bf 8} 1049-51   
    
\bibitem{gr} Grosjean C C 1987 The invariant density for a class of discrete-time maps involving an arbitrary monotonic function operator and an integer parameter {\it J. Math. Phys.} {\bf 28}(6) 1265-74
  1265-1274;

\bibitem{gth} Grossmann S and Thomae S 1977 Invariant Distributions and Stationary Correlation Functions of One-Dimensional Discrete Processes {\it Z. Naturforsch.} {\bf
    32a} 1353.

\bibitem{kch} Kuczma M, Chaczewski B and Ger R (eds) 1990 {\it Iterative Functional Equations} Encyclopedia of Mathematics and its Applications {\bf 32} (Cambridge: Cambridge University Press)

\bibitem{kcz} Kuczma M 1968 {\it Functional Equations in a single variable} (Warszawa: The
Polish Scientific Publishers)

\bibitem{kf} Katsura S and Fukuda W 1985 Exactly Solvable Models Showing Chaotic Behavior {\it Physica A} {\bf 130 A} 597-605 

\bibitem{kh} Katok A and Hasselblatt B 1995 {\it Introduction to  Modern Theory of Dynamical Systems} (Cambridge: Cambridge Univ. Press)

\bibitem{l} Lasota A 1973 Invariant Measures and Functional Equations {\it Aequationes Math.} {\bf 9} 193-200.

\bibitem{lmc} Lasota A and Mackay M C 1994 {\it Chaos, Fractals and Noise}  
  Appl. Math. Sc. {\bf 97} (New York: Springer)
 
\bibitem{m} Mackay M C 2003 {\it Time's Arrow} (New York: Dover Publications Inc.)

\bibitem{mil} Milnor J 2004 On Latt\`es Maps {\it Stony Brook IMS Preprint} {\bf \#2004/01} 

\bibitem{mr} Mira Ch 1982 Equation de Schr\"oder et solution des r\'ecurrences. G\'en\'eralization des polyn\^omes de Tchebysheff {\it Ed. du C.N.R.S., Tolouse} {\bf 332} 35-43 (in French)

\bibitem{sch} Schr\"oder E 1871 Ueber iterirte Functionen {\it Math. Ann.} {\bf 3} 296-322. Available through 
http://gdz.sub.uni-goettingen.de/en (in German, search by title)

\bibitem{sg} Skowroner L and Gora P 2007 Chaos in Newtonian Iterations: Searching for zeros which are not there {\it Acta Phys. Polonica} {\bf B 38}(5) 1909-24

\bibitem{sr} Souriac P 1982 R\'ecurrences avec Solutions Explicites, see \cite{mr} 69-73 (in French)

\bibitem{u1} Umeno K 1997 Method of constructing exactly solvable chaos {\it Phys. Rev. E} {\bf 55} 5280-84; Umeno K 1999 Exactly Solvable Chaos and Addition Theorems of Elliptic Functions {\it RIMS Kokyuroku}, {\bf 1098} 104-17 

\end{thebibliography}
\end{document}